\documentclass{lncse}
\usepackage{graphicx}

\begin{document}

\title{Order Parameter as an Additional State Variable of Unstable Traffic Flow}
\titlerunning{Order Parameter as State Variable}
\author{Ihor A. Lubashevsky\inst{1} \and Reinhard Mahnke\inst{2}}
\institute{Theory Department, General Physics Institute, Russian
Academy of Sciences,\\Vavilov str., 38, Moscow 117942, Russia
\and
Fachbereich Physik, Universit\"at Rostock, D--18051 Rostock, Germany}
\maketitle

\begin{abstract}
We discuss a phenomenological approach to the description of unstable 
vehicle motion on multilane highways that could explain in a simple way
such observed self-organizing phenomena as the sequence of the phase 
transitions ``free flow $\rightarrow$ synchronized motion $\rightarrow$ jam''
and the hysteresis in them.

We introduce a new variable called order
parameter that accounts for possible correlations in the vehicle motion 
at different lanes. So, it is principally due to ``many-body'' 
effects in the car interaction in contrast to such variables as the 
mean car density and velocity being actually the zeroth and 
first moments of the ``one-particle'' distribution function. Therefore, we
regard the order parameter as an additional independent state variable
of traffic flow and formulate the corresponding evolution equation 
governing the lane changing rate.

In this context we analyze the instability of homogeneous traffic flow 
manifesting itself in both of these phase
transitions and endowing them with the  hysteresis. Besides, the jam state
is characterized by the vehicle flows at different lanes being 
independent of one another.
\end{abstract}

\section{Introduction}

The existence of a new basic phase in vehicle flow on multilane
highways called the synchronized motion was recently discovered by 
Kerner and Rehborn \cite{KR2}, impacting significantly the physics
of traffics as a whole. In particular, it turns out that the 
spontaneous formation of moving jams on highways proceeds mainly
through a sequence of two transitions: ``free flow $\rightarrow$ 
synchronized motion $\rightarrow$ stop-and-go pattern'' \cite{K98}.  
Besides, all these transitions exhibit the hysteresis \cite{K98,KR1,KR3}.
As follows from the experimental data \cite{KR2,KR1,KR3} the
synchronized mode is essentially a multilane effect. Recently Kerner 
\cite{KL1,KL2} assumed that the transition ``free flow $\rightarrow$
synchronized mode'' is caused by ``Z''-like form of the overtaking
probability depending on the car density. 

There have been proposed several macroscopic models dealing with multilane 
traffic flow \cite{H97,HT98,HT99,HS99,Siam,L97,T99a,T99b}. Both these models 
specify the traffic dynamics completely in terms of the car density
$\rho$, mean velocity $v$, and, may be, the velocity 
variance $\theta$ or ascribe these quantities to the vehicle flow at each lane 
individually. Nevertheless, a quantitative description of the synchronized
mode is far from being developed well because of its complex structure
\cite{KL1,KL2}. In particular, it can form the totally homogeneous (\textit{i})
and homogeneous-in-speed (\textit{ii}) flows \cite{KR2}. 
Especially in the latter case there is no  
explicit relationship between the mean car velocity $v$ and density 
$\rho$, with the value of $v$ being actually constant and less then
that of free flow. The other important feature is 
the key role of some cars bunched together and traveling much faster 
than the typical ones, which enables to regard them as a special car
group \cite{KR2}. 
Therefore, in the synchronized mode the function of car distribution 
in the velocity space should have two maxima and we will call such fast car
groups platoons in speed. These features of the synchronized mode
have been substantiated also in \cite{Last} using single-car-data. In
particular, it has been demonstrated that the synchronized mode exhibits   
small correlations  between fluctuations in the car flow, velocity 
and density. There is only a strong correlation between the velocities at 
different lanes taken at the same time and decreasing sufficiently 
fast as the time difference increases. By contrast, there are strong long-time 
correlations between  the flow and density in the free flow state as well 
as the stop-and-go mode. 

Keeping in mind a certain analogy with aggregation processes in physical
systems Mahnke et al. \cite{M1,M2} proposed a kinetic model for the
formation of the
synchronized mode treated as the motion of a large car cluster. 
In the present paper following practically the spirit of the Landau theory
of phase transitions  we 
develop a phenomenological approach to the description of this process.
We ascribe to the vehicle flow an additional {\it internal} parameter
will be
called below the order parameter $h\in (0,1)$ characterizing the possible
correlations in the vehicle motion  at different lanes and write for it
a governing equation. For the car motion
where drivers do not change lane at all we set $h=0$, in the opposite limit
$h=1$. 

\section{Order Parameter and the Individual Driver Behavior\label{sec:2}}

For fixed values of $\rho$ and  $v$ the order parameter $h$ is assumed to be 
uniquely determined, thus, for a uniform vehicle flow we write: 
\begin{equation}
\tau \frac{dh}{dt}=-\Phi (h,\rho ,v)\enspace,  \label{2.1}
\end{equation}
where $\tau$ is the delay time and the function $\Phi(h,\rho,v)$ fulfills the 
inequality: 
\begin{equation}
\frac{\partial \Phi }{\partial h}>0\enspace.  \label{2.2}
\end{equation}
We note that the time $\tau$ characterizes the delay in the driver decision 
of changing lanes but not in the control over the headway, so, this delay
can 
be prolonged. The particular value $h(v,\rho )$ of the order parameter results 
from the compromise between the danger of an accident during changing lanes and 
the will of driver to move as fast as possible. Obviously, the lower is the mean 
vehicle 
velocity $v$ for a fixed value of $\rho$, the weaker is the lane-changing  
danger and the stronger is the will to move faster. Besides, the higher is 
the vehicle density $\rho $ for a fixed value of $v$, the stronger is this 
danger (here the will has no effect). Thus, the dependence $h(v,\rho)$
is an decreasing function of $v$ and $\rho $, so, due to (\ref{2.2}): 
\begin{equation}
\frac{\partial \Phi }{\partial v}>0\ ,\quad \frac{\partial \Phi }{\partial
\rho }>0\enspace ,  \label{2.3}
\end{equation}
with the latter inequality being caused by the danger effect only.
Equation (\ref{2.1}) describes actually the behavior of the drivers that
prefer to move faster than the statistically mean vehicle and whose
readiness for risk is greatest. Exactly this group of drivers
(platoons in speed) govern the value of $h$. 

There is, however, another characteristics of the driver behavior, it is 
the mean velocity $v=\vartheta (h,\rho )$ chosen by the \textit{statistically 
mean} driver taking into account also the danger resulting from the 
frequent lane changes by the ``fast'' drivers. Following typical assumptions
the velocity  $\vartheta (h,\rho )$ as a function of $\rho$ is considered 
to be decreasing:
\begin{equation}
\frac{\partial\vartheta }{\partial\rho} <0\quad \mathrm{and}\quad \rho \vartheta 
(\rho)\rightarrow 0\quad \mathrm{as}\quad \rho \rightarrow \rho _{0}\enspace ,
\label{intro:3}
\end{equation}
where $\rho _{0}$ is the upper limit vehicle density on road. In general,
the dependence of $\vartheta (h,\rho)$ on $h$ should be increasing for small 
values of the vehicle density, $\rho \ll \rho _{0}$, because in this case 
the lane-changing makes no substantial danger to traffic and practically all 
the drives can pass by vehicles moving at lower speed without risk. By contrast, 
when the vehicle density is sufficiently high, $\rho \sim \rho _{0}$, the 
lane-changing  is due to the car motion of the most ``impatient'' drivers 
whose behavior makes an additional danger to the main part of other drivers 
and the velocity $\vartheta (h,\rho )$ has to decrease as the order parameter $h$ 
increases. For certain intermediate values of the vehicle density, $\rho \approx 
\rho _{c}$, this dependence is to be weak as well as near the boundary
points, so: 
\begin{equation}
\frac{\partial \vartheta }{\partial h}>0\ \mathrm{for}\ \rho < \rho_c\,,\quad 
\frac{\partial \vartheta }{\partial h}<0\ \mathrm{for}\ \rho > \rho_c\,,\quad 
\frac{\partial \vartheta }{\partial h}=0\ \mathrm{at}\ h=0,1\enspace .
\label{2.3a}
\end{equation}
Then the governing equation (\ref{2.1}) takes the form:
\begin{equation}
\tau \frac{dh}{dt}=-\phi (h,\rho )\,,\ \mathrm{where}\ \phi (h,\rho ) 
\stackrel{\mathrm{def}}{=}\Phi [h,\rho ,\vartheta (h,\rho )] 
\label{2.4}
\end{equation}
and the condition $\phi(h,\rho )=0$ specifies the steady state dependence 
$h(\rho )$ of the order parameter on the vehicle density. 

Let us, now, study properties and stability of this steady state solution. 
From Eq.~(\ref{2.4}) we get
\begin{equation}
\frac{\partial \phi }{\partial h} = \frac{\partial \Phi }{\partial h}+ 
\frac{\partial \Phi }{\partial v}\,\frac{\partial \vartheta }{\partial h}\ ,
\qquad
\frac{\partial \phi }{\partial \rho } = \frac{\partial \Phi }{\partial \rho}
+\frac{\partial \Phi }{\partial v}\,\frac{\partial \vartheta }{\partial \rho }
\enspace .  \label{2.7}
\end{equation}
As mentioned above, the value of $\partial \Phi /\partial \rho $ is solely
due to the danger during changing lanes, so this term can be ignored until
the vehicle density $\rho $ becomes sufficiently high. Thus, in a 
certain region $\rho <\rho_h< \rho _{0}$ the derivative $\partial \phi
/\partial \rho \sim (\partial \Phi /\partial v)(\partial \vartheta /\partial
\rho )<0$ by virtue of (\ref{2.3})  and (\ref{intro:3})  and 
the function $h(\rho )$ is increasing or decreasing  for 
$\partial \phi /\partial h>0$ or $\partial \phi /\partial h<0$, respectively.
This statement follows directly from the relation 
$d h/d\rho = - \left(\partial \phi/\partial \rho\right)
\left(\partial \phi/\partial h\right)^{-1}$.

\begin{figure}[tbp]
\begin{center}
\includegraphics[scale=0.75]{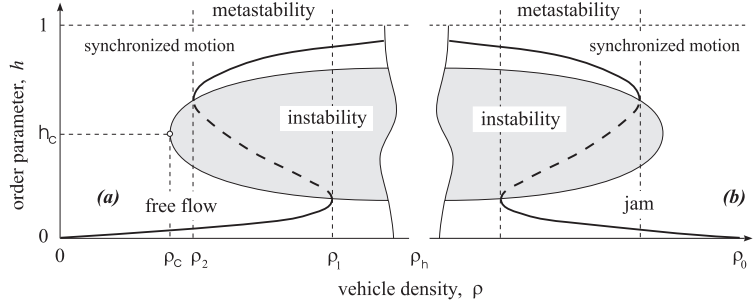}
\end{center}\vspace*{-\baselineskip}
\caption{The region of the traffic flow instability in the $h\rho$-plane and
the form of the curve $h(\rho)$ displaying the dependence of the order parameter 
on the vehicle density.}
\label{F6}
\end{figure}

For long-wave perturbations of the vehicle distribution on a highway the 
density $\rho $ can be treated as a constant. So, according to the 
governing equation~(\ref{2.4}), the steady-state traffic flow is unstable 
if $\partial \phi /\partial h<0$.
Due to (\ref{2.2}) and (\ref{2.3a}) the first term in the expression for 
$\partial \phi/\partial h$ in (\ref{2.7}) is dominant in the vicinity of the 
lines $h=0$ and $h=1$, thus, in these regions the curve $h(\rho )$ is 
increasing and the steady  state traffic flow is stable. For $\rho <\rho _{c}$ 
the value $\partial\vartheta /\partial h >0$, inequality~(\ref{2.3a}), and, 
thereby, 
the region $\left\{ 0<h<1,\;0<\rho <\rho _{c}\right\} $ corresponds to the stable
vehicle motion. However, for $\rho >\rho _{c}$ there can be an interval of the
order parameter $h$ where the derivative $\partial \phi /\partial h$ changes
the sign and the vehicle motion becomes unstable. Therefore, as the car density
$\rho$ grows causing the increase of the order parameter $h$ it can go into
the instability region wherein $dh/d\rho <0$. 
Under these conditions the curve $h(\rho)$ is to look like ``S'' 
(Fig.~\ref{F6}a) 
and its decreasing branch corresponds to the unstable 
vehicle flow. The lower increasing branch matches the free-flow state,
whereas the upper one should be related to the synchronized phase because 
it is characterized by the order parameter coming to unity.

\section{Phase Transitions and the Fundamental Diagram}

The obtained dependence $h(\rho)$ actually describes the first order phase 
transition in the vehicle motion. Indeed, when increasing the car density 
exceeds the value $\rho_1$ the free flow becomes absolutely unstable and
the synchronized mode forms through a sharp jump in the order parameter. If,
however, after that the car density decreases  the synchronized mode will persist
until the car density attains the value $\rho_2 < \rho_1$. It is a typical
hysteresis and the region $(\rho_2, \rho_1)$ corresponds to the metastable
phases of traffic flow. It should be noted that the stated approach to the 
description of the phase transition ``free flow $\rightarrow$
synchronized mode'' is rather similar to the hypothesis by Kerner
\cite{KL1,KL2} about ``Z''-like dependence of the overtaking
probability on the car density that can cause this phase transition.   

\begin{figure}[tbp]
\begin{center}
\includegraphics[scale=0.75]{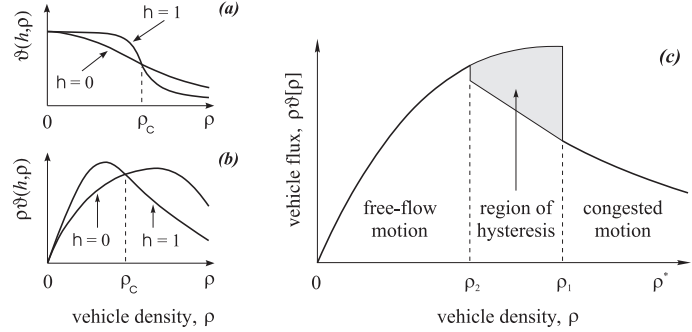}
\end{center}\vspace*{-\baselineskip}
\caption{The mean vehicle velocity ({\protect\boldmath $a$}) and the vehicle
flux ({\protect\boldmath $b$}) vs. the vehicle density for the limit values
of the order parameter $h=0,1$ as well as the resulting fundamental diagram 
({\protect\boldmath $c$}).}
\label{F7}
\end{figure}

Let us, now, discuss a possible form of the fundamental diagram showing 
$j=\rho \vartheta [\rho ]$  where, by
definition, $\vartheta [\rho ]=\vartheta [h(\rho ),\rho ]$. 
Fig.~\ref{F7}a displays the dependence $\vartheta (h,\rho )$ of the
mean vehicle velocity on the density $\rho $ for the fixed limit values of
the order parameter $h=0$ or 1. For small values of $\rho $ these curves
practically coincide with each other. As the vehicle density $\rho $ grows
and until it comes close to the critical value $\rho _{c}$ where the lane
change danger becomes substantial, the velocity $\vartheta (1,\rho )$
practically does not depend on $\rho $. So at the point $\rho _{c}$ at which
the curves $\vartheta (1,\rho )$ and $\vartheta (0,\rho )$ meet each other
the former curve, $\vartheta (1,\rho )$, is to exhibit sufficiently sharp
decrease in comparison with the latter one. Therefore, on one hand, the
function $j_{1}(\rho )=\rho \vartheta (1,\rho )$ has to be decreasing for 
$\rho >\rho _{c}$. On the other hand, at the point $\rho _{c}$ for $h\ll 1$
the effect of the lane change danger is not extremely strong, it only makes
the lane change ineffective, $\partial \vartheta /\partial h\approx 0$
(compare (\ref{2.3a})). So it is reasonable to assume the function $
j_{0}(\rho )=\rho \vartheta (0,\rho )$ increasing neat the point $\rho _{c}$.
Under the adopted assumptions the relative arrangement of the curves $
j_{0}(\rho )$, $j_{1}(\rho )$ is demonstrated in 
Fig.~\ref{F7}b, and Fig.~\ref{F7}c shows the
fundamental diagram of traffic flow resulting from Fig.~\ref{F6}
and Fig.~\ref{F7}b.

The developed model predicts also  the same type phase transition for large
values of the order parameter.
In fact, in an extremely dense traffic flow changing
lanes is sufficiently dangerous and the function $\Phi (h,v,\rho )$
describing the driver behavior is to depend strongly on the vehicle density 
as $\rho \rightarrow \rho _{0}$. In addition, the vehicle motion
becomes slow. Under such conditions the former term in the expression for 
$\partial \Phi/\partial\rho$ in (\ref{2.7}) should be dominant and, so,
$\partial \phi/\partial \rho >0$ and the
stable vehicle motion corresponding to $\partial \phi /\partial h>0$ is
characterized by the decreasing dependence of the order parameter $h(\rho )$
on the vehicle density $\rho $ for $\rho >\rho_h$. Therefore, as the vehicle
density $\rho $ increases the curve $h(\rho )$ can again go into the instability
region (in the $h\rho $-plane), which has to give rise to a jump
from the synchronized mode to a jam. The latter matches
small values of the order parameter $h$ 
(Fig.~\ref{F6}b), so, it should comprise the vehicle flows along
different lane where lane changing is depressed, making them practically 
independent of one another.  

\section{Conclusion\label{sec:cr}}

We have introduced an additional state variable of the traffic flow,
the order parameter $h$,
that
accounts for internal correlations in the vehicle motion caused by the 
lane changing. Since such correlations are due to the ``many-body'' effects
in the car interaction the order parameter is regarded as an independent
state variable.
Keeping in mind general properties of the driver behavior
we have written the governing equation for this variable. 

It turns out that in this way such characteristic properties of the traffic flow
instability as the sequence of the phase transitions  
``free flow $\rightarrow$ synchronized motion $\rightarrow$ jam''
can be described without additional assumptions. Moreover, in this model
both the phase transitions are of the first order and exhibits
hysteresis. Besides, the synchronized mode corresponds to highly correlated
vehicle flows along different lanes, $h\approx 1$, whereas in the free
flow and the jam these correlations are depressed, $h \ll 1$. 
So, the jam phase actually comprises mutually independent car flows
along different lanes.

\end{document}